\documentclass[12pt,preprint]{aastex}
\begin{document}
\title{The Alignments of Disk Galaxies with the Local Pancakes} 
\author{Yookyung Noh and Jounghun Lee}
\affil{Department of Physics and Astronomy, Seoul National University, 
Seoul 151-742, Korea} 
\email{ykyung@astro.snu.ac.kr, jounghun@astro.snu.ac.kr}

%%%%%%%%%%%%%%%%%%%%%%%%%%%%%%%%%%%%%%%%%%%%%%%%%%%%%%%%%%%%%%%%%%%%%
\begin{abstract}
We analyze the Tully catalog of nearby galaxies to investigate the local 
pancaking effect on the orientation of disk galaxies. We first select only 
those edge-on disk galaxies in the catalog whose axis-ratios are less than 
$0.1$ to measure their spin axes unambiguously.  A local pancake at the 
location of each selected galaxy is found as a plane encompassing the two 
nearest neighbor disks.  Then, we examine statistically the inclinations of 
the galaxy spin axes relative to the local pancake planes. It is detected 
that the Tully disk galaxies tend to be inclined onto the local pancake 
planes, and the average inclination angles decrease with the pancake scale.
We also construct a theoretical model for the inclination of disk galaxies 
relative to the local pancakes in the frame of the linear tidal torque theory. 
The comparison of the theoretical prediction with the observational result 
demonstrates a good agreement.  Finally, we conclude that it is a first 
detection of the local pancaking effect on the orientation of disk galaxies, 
which is consistent with the scenario that the gravitational tidal field 
promotes the formation of pancakes on small mass scale. 
\end{abstract}
%%%%%%%%%%%%%%%%%%%%%%%%%%%%%%%%%%%%%%%%%%%%%%%%%%%%%%%%%%%%%%%%%%%%%%%

\section{INTRODUCTION}

The standard theory of structure formation based on the cold dark matter 
(CDM) paradigm explains that the building blocks of cosmic structures in the 
universe are virialized halos made up of cold dark matter particles, and that 
these CDM halos form hierarchically via gravity from the primordial 
density fluctuations with a characteristic spectrum.
 
Although this standard theory has been remarkably successful in explaining 
the observed properties of the universe on the galaxy cluster scale, 
its validity on the galactic scale is still inconclusive, since observational 
data seemed inconsistent with the theory on that scale \citep[e.g.,][]
{kau-etal93,kly-etal99,moo-etal99,bin-eva01}.

Confronted with observational challenges, various theoretical attempts 
have been recently made to refine the standard theory of structure formation 
on the galactic scale.  One of such attempts is the currently proposed  
``broken hierarchy'' scenario \citep[e.g.,][and references therein]
{bow-etal05}.  
According to this idea, the gravitational clustering to form dark halos does 
not proceed in a perfectly hierarchical way on all scales, unlike the
prediction of the standard theory. Rather, the scenario suggests that 
the clustering process take on a somewhat anti-hierarchy on small mass scales 
as the gravitational {\it tidal} field favors the formation of a pancake 
(an object collapsed along one direction) over that of a halo (an object 
collapsed along all three directions).  This scenario has been shown to be 
capable of resolving several conflicts with observational data. 
For example, \citet{mo-etal05} recently showed that if the pre-virialization 
induced by the pancaking effect is taken into account, 
the observed HI mass function and the faint-end slope 
of the galaxy luminosity function can be explained within the context of 
the standard structure formation theory. 

Promising as this scenario seems, it is not an easy task to test it against 
observations and to detect directly the local pancaking effect.  
A possible way to detect the pancaking effect is to investigate the 
{\it anisotropy} in the orientation of galaxies induced by the gravitational 
tidal field.  According to the tidal torque theory, the angular momentum 
of a galaxy is originated by the tidal interaction with the surrounding 
matter. A generic prediction of the tidal torque theory is that the direction 
of the galaxy angular momentum is not random but aligned with a local tidal 
shear tensor \citep[e.g.,][]{dub92,lee-pen00}.  
Thus, if a galactic halo is embedded in a local pancake that formed earlier 
than the galactic dark halo, then a correlation between the galaxy spin axis 
and the pancake principal axis should exist. 

In fact, it was claimed by many observational reports that the galaxy 
spin axes are correlated with the surrounding matter distribution
\citep{hel-sal82,fli-god86,kas-oka92,gar-etal93}. Very recently, more 
convincing evidences for the anisotropy in the orientations of the galaxy spin 
axes induced by the tidal fields of the surrounding matter have been found.  
\citet{nav-etal04} detected that the spin axes of the nearby edge-on spirals 
in the Principal Galaxy Catalog \citep[PGC,][]{pat-etal97} have a strong 
tendency to lie parallel to the Super-galactic plane (SGP). 
\citet{tru-etal05} also found that the spin axes of the galaxies located near
the shells of the largest cosmic voids lie preferentially on the void surface. 
 
Note, however, that these observations found evidences only for those 
particular cases that the galaxies are surrounded by the large-scale structure 
with a flat surface.  In other words, what was detected so far is not the 
effect of the {\it local} pancake but the effect of the {\it large scale} 
coherence of the surrounding matter on the orientations of the galaxy spin 
axes. Here, our goal is to detect a local pancaking effect by investigating 
the anisotropy on the orientation of the galaxy spins from observational 
data. In $\S 2$, we determine the probability density distribution of 
the inclination angles of the Tully disk galaxies relative to the local 
pancake planes. In $\S 3$, we model analytically the anisotropy in the 
orientation of the galaxy spin axes in the frame of the tidal torque theory, 
and compare the analytic prediction with observational data. 
The results are discussed and a final conclusion is drawn in $\S 4$.

\section{OBSERVATIONAL ANALYSIS}

To detect the intrinsic alignment of the galaxies and local pancake 
in the real universe, we use the Tully catalog. In the Tully catalog, 
$35674$ nearby galaxy properties observed from the whole sky are piled up. 
Among them we make a subsample of $12122$ disk galaxies which have a type 
$0-9$ listed in Third Reference Catalog of Bright Galaxies 
\citep[RC3,][]{vau91}.  The median redshift of the subsample is 
$5180$ km/s. Note that the gravitational weak lensing effect can be 
neglected at this low redshift \citep{man-etal05}.

We focus our analysis on this subsample of spirals for the following reason.
The spiral galaxies have relatively low velocity dispersion, approximately 
$126\pm10$ km/s \citep{dav-etal97}.  It indicates that the spiral 
galaxies did not move far from their initial Lagrangian positions in the 
subsequent evolution. Thus, the positions of the spirals may be optimal to 
define the local pancake planes.

First, we measure the spin direction of the spiral galaxies by using 
the position angle (PA) on the sky and the projected axial ratio ($e$). 
Assuming that the geometry of the sky is flat, the spin vector of each galaxy
in the spherical coordinate system is obtained from $\tan(\textrm{PA})
=\hat{L_{\theta}}/\hat{L_{\phi}}$ and $e=\vert\hat{L_{r}}\vert$. For the
calculation of the spin vectors,  two fold degeneracy should be accounted for. 
In other words, it is hard to decide whether the galaxy rotates clockwise
$(\hat{L_{r}})$ or anticlockwise $(-\hat{L_{r}})$. We consider the 
degeneracy problem statistically as follow. The probability distribution 
$p(\cos\theta)$ is calculated by $ [p(\cos\theta_{1}) + 
p(\cos\theta_{2})]/2$, where $\theta_{1}$ and $\theta_{2}$ are the angles 
of the unit vector normal to the local pancake with $\hat{L_{r}}$ and 
$-\hat{L_{r}}$, respectively. To reduce the degeneracy, we use only edge-on 
galaxies whose axial ratio is less than 0.1. The total number of the edge-on 
galaxies is found to be $116$. The spin directions calculated in the 
spherical coordinates are transformed to those in the Cartesian coordinates 
with the given equatorial coordinate positions (1950 equinox for RA and DEC).

\begin{figure} \begin{center} \plotone{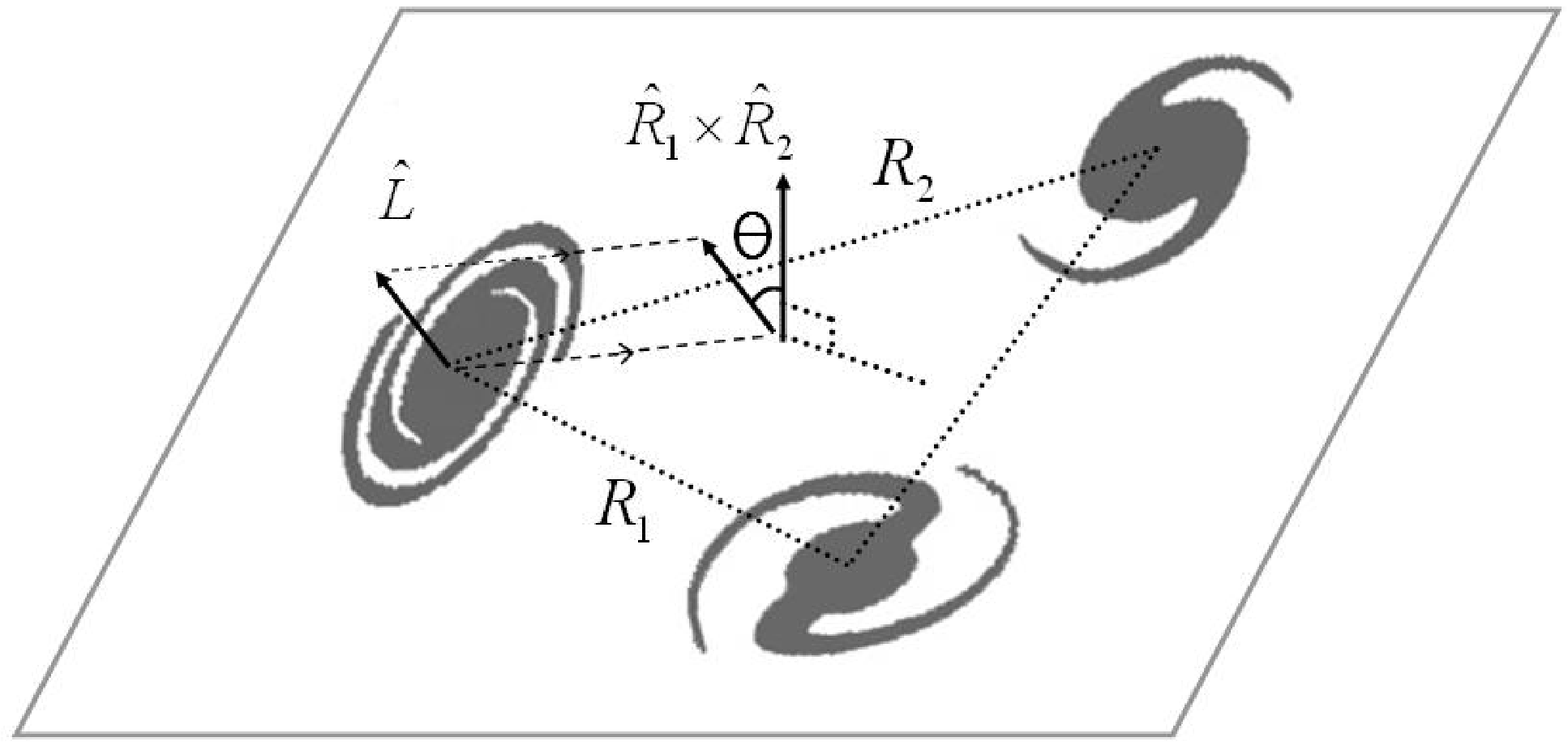}
\caption{The illustration of a Tully disk galaxy embedded in a local pancake. 
$\bf{\hat L}$ represents a spin vector of a galaxy. $\bf{R_{1}}$ and 
$\bf{R_{2}}$ are the distances to the two nearest neighbors, respectively. 
$\bf{R_{1}}\times \bf{R_{2}}$ corresponds to  a unit normal vector of the 
local pancake plane. $\theta$ represents the angle between the spin vector 
and the normal vector of the plane.}
\label{fig:plane} \end{center} \end{figure}

Now, a local pancake has to be determined at each given edge-on spiral whose 
spin vector is obtained. As mentioned in $\S 1$, theoretically the local 
pancake is defined as a sheet-like structure gravitationally collapsed 
along the major principal axis of a local shear. In practice, however, 
it is hard to apply this theoretical method for the observational data.
Instead, we employ the  following practical scheme. At the location of 
each edge-on spiral, we first find the two nearest neighbors among all $12122$ 
spirals in the subsample, and determine a local pancake as the plane 
encompassing the galaxy itself and its two nearest neighbors. 
Figure \ref{fig:plane} illustrates the configuration of an edge-on galaxy 
and its local pancake.  The logic for this scheme is as follows: 
If the formation of a small-mass galactic halo is really preceded by that of 
a larger-mass local pancake at the early epoch, then the two-dimensional 
coherence in the locations of the neighbor galaxies should be expected. 
In other words, the placements of the nearest neighbors should be more or 
less arranged in the plane of the local pancake where they are all embedded. 
The scale of a local pancake at a given edge-on spiral is determined by the 
distance from the galaxy to the second nearest neighbor.  
It is worth mentioning again that we consider only the spiral galaxies 
since their peculiar velocities are relatively small so that they
are expected to conserve their initial Lagrangian position.

\begin{figure} \begin{center} \plotone{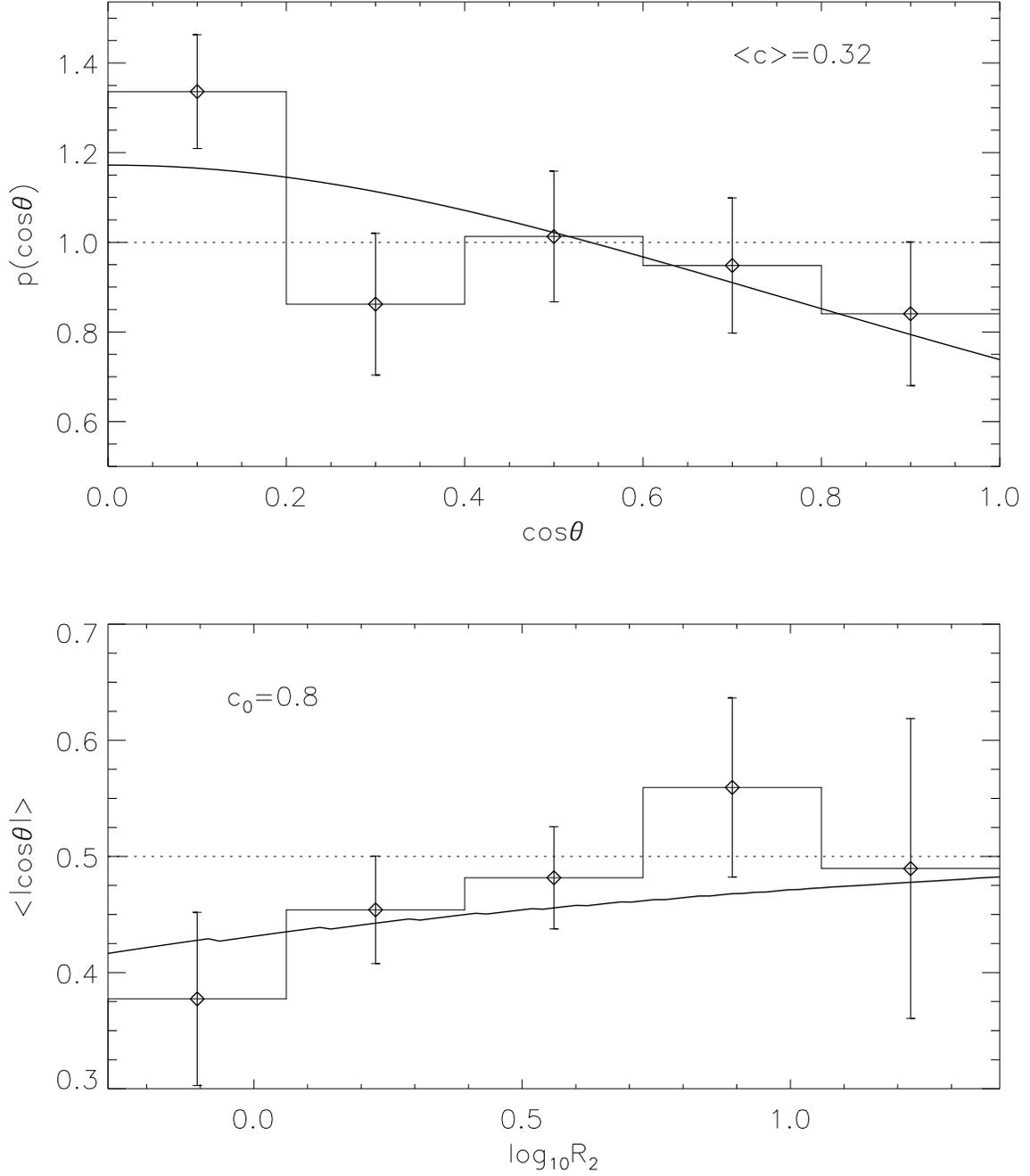}
\caption{{\it Upper} : the probability density distribution of the cosines 
of the angles between the galaxy spin axis and the unit vector normal to 
the local pancake plane. The observational data are plotted as squre dots 
with Poisson error bars. The analytic prediction (eq. [\ref{eq:Ptheta}]) 
is also plotted as solid line with the best-fit mean value of 
$\langle c\rangle = 0.32$.  The horizontal dotted line corresponds to the 
case that there is no alignment. 
{\it Lower} : The average value of the cosines of the angles between 
the galaxy spin axis and the unit vector normal to the local pancake plane
as a function of the pancake scale. The squares with the histogram 
represent the observational result with the Poisson errors. 
The theoretical prediction (eq. [\ref{eq:corr}]) with $c_{0}=0.8$ is plotted 
as a solid line.  The horizontal dotted line corresponds to the 
case that there is no alignment.}
\label{fig:prob} \end{center} \end{figure}

Now that a local pancake at the location of each edge-on disk is defined,  
we compute the cosines of the angles between the spin axis and the unit 
vector normal to the pancake plane (see Fig. \ref{fig:plane}) as 
$\cos\theta=\vert{\bf\hat{L}}\cdot({\bf\hat{R_{1}}}\times{\bf\hat{R_{2}}})
\vert$, where $\theta\in[0,\pi/2]$, ${\bf\hat{L}}$ is the unit spin vector 
of an edge-on galaxy, and ${\bf\hat{R_{1}}}, {\bf\hat{R_{2}}}$ are 
the unit vectors of the displacement to the first and the second nearest 
neighbors, respectively. The unit vector normal to the local pancake is 
determined as ${\bf\hat{R_{1}}}\times{\bf\hat{R_{2}}}$. From this, we compute 
the probability density distribution, $p(\cos\theta)$, and the average, 
$\vert\cos\alpha\vert$, which are shown as histograms with the error bars 
in the upper and the lower panels of Figure \ref{fig:prob}, respectively.  
The errors for estimating $p(\cos\theta)$ are just Poissonian.
For the errors involved in measuring $\langle|\cos\theta|\rangle$, we 
use the statistical deviation, $\sqrt{(\frac{1}{3}-{\frac{1}{2}}^{2})/N}$,  
($N$: the number of galaxies in each bin) from the no-inclination case  
of $\langle\vert\cos\theta\vert\rangle = 1/2$ and 
$\langle\vert\cos^{2}\theta\vert\rangle = 1/3$. 

The observational results shown in Figure \ref{fig:prob} reveal that  
the spin axes of the disk galaxies tend to be inclined onto the plane of the 
local pancakes, and that the degree of the inclination decreases with the 
pancake scale. It reflects that the galaxy inclination to the pancake 
plane is indeed a {\it local} correlation with the gravitational tidal field.  
Suffering from the large error-bars due to the small sample size, we test the 
rejection of the no-inclination hypothesis, using the observed data points. 
It is found that the no-inclination hypothesis is rejected at the $89\%$ 
confidence level.  

\section{THEORETICAL ANALYSIS}

To construct a theoretical model for the inclinations of disk galaxies 
relative to the local pancakes, we assume the following.
\begin{enumerate}
\item
The angular momentum of a disk galaxy is originated by the initial tidal 
interaction with the surrounding matter \citep{dor70, whi84}, which induces 
a correlation in the orientation of the galaxy angular momentum vector, 
${\bf L}=(L_{i})$, with the local tidal shear tensor, ${\bf T}=(T_{ij})$ 
defined as the second derivative of the perturbation potential. 
\item
On small-mass halo scales, the gravitational tidal field promotes the 
formation of a pancake \citep{zel70,mo-etal05}
that precedes  the formation of a low-mass halo. The formation of a local 
pancake is well described by the Zel'dovich approximation \citep{zel70}, 
according to which the local density at present epoch is given as:
\begin{equation}
\label{eq:zel}
\rho = \frac{\bar{\rho}}{(1-\lambda_{1})(1-\lambda_{2})(1-\lambda_{3})},
\end{equation}
where $\bar{\rho}$ is the mean density of the Universe, $\lambda_{1},
\lambda_{2},\lambda_{3}$ are the three eigenvalues of the local tidal shear 
tensor, ${\bf T}$ in a decreasing order. Equation (\ref{eq:zel}) implies 
that the formation of a local pancake (i.e., a first shell crossing) occurs 
at present epoch if the largest eigenvalue, $\lambda_{1}$, reaches unity.
\item
The minor principal axis of the inertia momentum tensor of a local pancake 
is aligned with the major principal axis of the local shear tensor, since 
the gravitational collapse along the major principal axis of the local 
shear tensor forms a pancake. 
\item
Let ${\bf L} = (L\sin\theta\cos\phi,L\sin\theta\sin\phi,L\cos\theta)$ be 
a galaxy angular momentum vector in the principal axis system of the inertia 
momentum tensor of a local pancake. Now that \citet{lee-etal05} already found 
an analytic expression for the probability density distribution of the 
cosines of the angles between ${\bf L}$ and the major principal axis of 
the inertia momentum tensor (which is the minor principal axis of ${\bf T}$ 
according the above third assumption),  one can derive straightforwardly 
the probability density distribution of the cosines of the angles between 
${\bf L}$ and the {\it minor} principal axes of the inertia momentum tensors 
(which is the major principal axes of ${\bf T}$) by rotating the minor to the 
major principal axis as 
\begin{equation}
\label{eq:Ptheta}
p(\cos\theta) =
\frac{1}{2\pi}\prod_{i=1}^{3}(1+c-3c\hat{\lambda}^{2}_{i})^{-\frac{1}{2}}
\int_{0}^{2\pi}(\frac{\sin^{2}\theta\cos^{2}\phi}{1+c-3c\hat{\lambda}^{2}_{3}}+
\frac{\sin^{2}\theta\sin^{2}\phi}{1+c-3c\hat{\lambda}^{2}_{2}}+
\frac{\cos^{2}\theta}{1+c-3c\hat{\lambda}^{2}_{1}})
^{-\frac{3}{2}}d\phi,
\end{equation}
where, $c$ is a correlation parameter in the range of $[0,1]$. Here, 
$\{\hat{\lambda}_{1},\hat{\lambda}_{2},\hat{\lambda}_{3}\}$  represent the 
three eigenvalues (in a decreasing order) of the unit traceless tidal shear 
tensor, $\hat{\bf T} = (\hat{T}_{ij})$ defined as 
$\hat{T}_{ij}\equiv \tilde{T}_{ij}/|\tilde{\bf T}|$ with 
$\tilde{T}_{ij} \equiv T_{ij} - {\rm Tr}({\bf T})\delta_{ij}/3$, 
related to $\{\lambda_{1},\lambda_{2},\lambda_{3}\}$, as 
\begin{eqnarray}
\label{eqn:hlam1}
\hat{\lambda}_{1} &=& \frac{2\lambda_{1}-\lambda_{2}-\lambda_{3}}
{\sqrt{6(\lambda^{2}_{1}+\lambda^{2}_{2}+\lambda^{2}_{3}
-\lambda_{1}\lambda_{2} - \lambda_{2}\lambda_{3} - \lambda_{1}\lambda_{3})}},
\\
\label{eqn:hlam2}
\hat{\lambda}_{2} &=& \frac{-\lambda_{1}+2\lambda_{2}-\lambda_{3}}
{\sqrt{6(\lambda^{2}_{1}+\lambda^{2}_{2}+\lambda^{2}_{3}
-\lambda_{1}\lambda_{2} - \lambda_{2}\lambda_{3} - \lambda_{1}\lambda_{3})}},
\\
\label{eqn:hlam3}
\hat{\lambda}_{3} &=& \frac{-\lambda_{1}-\lambda_{2}+2\lambda_{3}}
{\sqrt{6(\lambda^{2}_{1}+\lambda^{2}_{2}+\lambda^{2}_{3}
-\lambda_{1}\lambda_{2} - \lambda_{2}\lambda_{3} - \lambda_{1}\lambda_{3})}}.
\end{eqnarray}
Basically, equation (\ref{eq:Ptheta}) is obtained from the formula given 
in Lee et al. (2005) simply by exchanging $\hat{\lambda}_{1}$ with 
$\hat{\lambda}_{3}$.
\item
The correlation parameter, $c$,  in equation (\ref{eq:Ptheta}) depends on 
the filtering scales on which the galaxy angular momentum ${\bf L}$ and the 
local shear tensor ${\bf T}$ are smoothed. The correlation parameter $c$ 
is a constant only if the two scales are the same. If they are different, 
then the correlation parameter is no longer a constant but 
varies with the filter scales \citep{lee-pen01} as
\begin{equation}
\label{eq:corr}
c = c_{0}\frac{\sigma(R_{2})}{\sigma(R_{G})}, 
\end{equation}
where $c_{0}$ is a constant, and $\sigma(R_{2})$ is a rms fluctuation of the 
linear density field smoothed on the pancake scale $R_{2}$. Here, $R_{G}$ 
represents the galactic scale $\approx 0.55 h^{-1}$Mpc \citep{bbks} 
on which the angular momentum vector ${\bf L}$ is smoothed. 
\end{enumerate}

To complete equation (\ref{eq:Ptheta}), one has to specify the values of 
$\{\hat{\lambda}_{i}\}_{i=1}^{3}$ and $c$. According to the second 
assumption, a local pancake of the scale length $R_{2}$ forms when the largest 
eigenvalue of $\lambda_{1}$ of the tidal shear tensor smoothed on the 
scale $R_{2}$ reaches unity. As for the other two eigenvalues, $\lambda_{2}$ 
and $\lambda_{3}$, one may use the most probable values under the constraint 
of $\lambda_{1}=1$.  
 
Using the probability density distributions, 
$p(\lambda_{1},\lambda_{2},\lambda_{3})$ and $p(\lambda_{1})$, derived by 
\citet{dor70}, and with the help of the Bayes theorem, we evaluate the 
constrained joint probability density distribution of $\lambda_{2}$ and 
$\lambda_{3}$ as 
\begin{equation}
\label{eq:Plambda}
p(\lambda_{2},\lambda_{3}\vert\lambda_{1}=1) = p(\lambda_{1}=1,\lambda_{2},
\lambda_{3})/p(\lambda_{1}=1).
\end{equation}
By equation (\ref{eq:Plambda}), we determine the most 
probable values of $\lambda_{2}$ and $\lambda_{3}$, which turn out to be  
almost scale-independent, changing only very mildly in the range of 
$[0.38,0.48]$ and $[0,0.1]$, respectively as the scale $R_{2}$ varies 
from $R_{G}$ to over $10 h^{-1}$ Mpc. The values of 
$\{\hat{\lambda}_{i}\}_{i=1}^{3}$ can be obtained straightforwardly 
through equations (\ref{eqn:hlam1})-(\ref{eqn:hlam3}). 

Now, we fit the observational data points obtained in $\S 2$ to equation 
(\ref{eq:Ptheta}) with the value $c$ as an adjustable free parameter.
The best-fit value of $c$ is found to be $\langle c \rangle \approx 0.32$ 
through the $\chi^{2}$ minimization. Note here that this best-fit value 
corresponds not to the constant $c_{0}$ in equation (\ref{eq:corr}) but 
to the {\it mean} value averaged over $R_{2}$. We also perform an analytic 
evaluation of the average value, 
$\langle\vert\cos\theta\vert\rangle$ as
\begin{equation}
\label{eq:Mval}
\langle\vert\cos\theta\vert\rangle = \int_{0}^{\pi/2}\vert\cos\theta\vert 
p(\cos\theta)d\theta. 
\end{equation}
Now that this average value, $\langle\vert\cos\theta\vert\rangle$, 
is a function of $R_{2}$ as the correlation parameter, $c$, in $p(\cos\theta)$ 
depends on $R_{2}$,  we can determine the best-fit value of the constant 
$c_{0}$ by fitting the observational data obtained in section $\S 2$ 
to equation (\ref{eq:Mval}) with equation (\ref{eq:corr}). We find 
$c_{0}\approx 0.8$. 

Figure \ref{fig:prob} depicts the analytically derived probability density 
distribution $p(\cos\theta)$ and the average value 
$\langle\vert\cos\theta\vert\rangle$ as solid lines in the upper and the 
lower panels, respectively. The comparison with these analytic results 
with the observational points reveals a fairly good agreement between the 
two. 

\section{DISCUSSION AND CONCLUSION}

We have presented an observational evidence for the preferential inclinations  
of the spin axes of the Tully disk galaxies onto the local pancake planes, 
and provided a quantitative theoretical explanation to the observed phenomena 
in terms of the tidal interaction and the broken hierarchy.  

Successful as the match between the theory and the observation seems, 
some limitations of our analysis deserve discussing here. First, in the 
observational analysis the redshift distortion effect is not properly taken 
into account. Even though the redshift distortion effect is unlikely to 
reduce the observed inclination strength, as \citet{tru-etal05} noted, 
it should be necessary to account for it to find a true signal. 
Second, in the theoretical analysis, we describe the formation of a local 
pancake in terms of the Zel'dovich collapse condition. 
This is an obvious oversimplification of the reality, as \citet{she-etal05} 
noted that the real collapse condition should be more complicated. 
Although the Zel'dovich collapse condition is a good approximation, 
a more realistic description of the formation of a local pancake would 
be desirable. 

Finally, we conclude that our observational and theoretical study 
of the local pancaking effect on the orientation of the galaxy spin axes 
will provide a new hint to the unsolved problem of the galaxy 
formation. 

\acknowledgments

This work is supported by the research grant No. R01-2005-000-10610-0 from the 
Basic Research Program of the Korea Science and Engineering Foundation.
We also acknowledge a partial support of the Brain Korea 21 Project in 2005.

\clearpage
%%%%%%%%%%%%%%%
%%%%%%%%%%%%%%%%

\clearpage
%%%%%%%%%%%%%%%
\end{document}